\documentclass[10pt]{article}
\usepackage[letterpaper]{geometry}
\usepackage{hicss}
\usepackage{times}
\usepackage[none]{hyphenat}
\usepackage{url}
\usepackage{latexsym}
\usepackage{indentfirst}
\usepackage{graphicx}
\graphicspath{{images/}}
\usepackage{hyperref}
\usepackage{cleveref}
\usepackage[square,numbers,sort]{natbib}

\title{Intelligent Decision Assistance Versus Automated Decision-Making: Enhancing Knowledge Work Through Explainable Artificial Intelligence}

 \author{Max Schemmer \\
  Karlsruhe Institute of Technology \\
  {\underline{max.schemmer@kit.edu}} \\\And
  Niklas Kühl \\
  Karlsruhe Institute of Technology \\
  {\underline{niklas.kuehl@kit.edu} }\\\And
  Gerhard Satzger \\
  Karlsruhe Institute of Technology \\
  {\underline{gerhard.satzger@kit.edu}} }

\date{}

\begin{document}
\maketitle
\begin{abstract}
%While recent advances in AI-based automated decision-making have shown many benefits for businesses and society, they also come at a cost. It has for long been known that a high level of automation of decisions can lead to various drawbacks, such as automation bias and deskilling. In particular, the deskilling of knowledge workers is a major issue, as they are the same people who should also train, challenge and evolve AI. To address this issue, we conceptualize a new class of DSS, namely Intelligent Decision Assistance (IDA) based on a literature review of two different research streams---Decision Support Systems (DSS) and automation. IDA supports knowledge workers without influencing them through automated decision-making. Specifically, we propose to use techniques of Explainable AI (XAI) while withholding concrete AI recommendations. To test this conceptualization, we develop hypotheses on the impacts of IDA and provide first evidence for their validity based on empirical studies in the literature. Our results confirm the validity of the conceptualization and motivate further research.
While recent advances in AI-based automated decision-making have shown many benefits for businesses and society, they also come at a cost. It has for long been known that a high level of automation of decisions can lead to various drawbacks, such as automation bias and deskilling. In particular, the deskilling of knowledge workers is a major issue, as they are the same people who should also train, challenge and evolve AI.
To address this issue, we conceptualize a new class of DSS, namely Intelligent Decision Assistance (IDA) based on a literature review of two different research streams---DSS and automation. IDA supports knowledge workers without influencing them through automated decision-making. Specifically, we propose to use techniques of Explainable AI (XAI) while withholding concrete AI recommendations. To test this conceptualization, we develop hypotheses on the impacts of IDA and provide first evidence for their validity based on empirical studies in the literature. 
%Optional: Motivate through synthesize of DSS and automation and take best of both worlds
\end{abstract}

\section{Introduction}
%Challenge how to make it interesting beside the idea --> Validate the idea through a meta study
%General LR --> Conceptualisation --> meta study
%TODO: Focus AI-based
%Intelligent decision Assistance 
%More tasks automated including knowledge work because of AI-> disadvantages of automation, deskilling --> an issue for knowledge work --> DSS literature (informative decisional guidance) --> solution --> verify and discuss solution

%Kernel theory is informative guidance
%TODOS: 
    %Check if we can quote other paper from track
    %Where do we position AI and XAI?

%What has changed
The recent advances in Artificial Intelligence (AI) lead to an increase in automated decision-making \cite{coombs2020strategic}. Decisions can be classified into unstructured, semi-structured, and structured decisions  \cite{turban2010decision}.
Traditionally, automated decision-making was applied to structured problems, and Decision Support Systems (DSS) enhanced decision-making for unstructured problems \cite{gorry1971framework,turban2010decision}.  
Unstructured tasks were considered too difficult to automate since they require more cognitive flexibility \cite{lacity2016new}.
However, advances in AI, specifically in deep learning, now increasingly enable to automate also more complex cognitive tasks, such as driving a car \cite{frey2017future}. 
Therefore, AI has now the potential to also address semi-structured and unstructured decisions that are far from basic back-office tasks \cite{asatiani2019implementation}.
For example, AI is used to automate loan approval \cite{Infosys2019}, or to conduct recruitment choices \cite{albert2019ai}---decisions that in the past were unimaginable to automate. 
Therefore, both the number and complexity of tasks that can be automated increase.

%Disadvantages 
However, it has long been known that increasing automation of decisions can lead to various drawbacks, such as automation bias and deskilling \cite{Mosier1999,Parasuraman2000}. 
This is especially challenging since most semi-structured and unstructured tasks are knowledge work incorporating high-stake decision making, e.g. medical diagnosis or jurisdictional decisions.
In general, AI for knowledge workers should automate routine and assist knowledge-intensive work with reasoning and other high-level functions \cite{adelstein2007disconnecting}.
The deskilling of knowledge workers is a major problem, as they are the people who should train, challenge and evolve AI. Knowledge workers create the labels for the AI that is the foundation for its initial training. 
After changes in the environment of the AI
knowledge workers adapt and develop new solutions based on their domain expertise \cite{baier2019cope}. Furthermore, they should be able to challenge the AI's recommendation, either with regard to performance but also with respect to ethical and fairness concerns. While in many use cases these disadvantages may be negligible there are cases where they must not be ignored. Reasons include, but are not limited to, losing significant competitiveness, e.g. in asset investment strategy decisions, or even potentially harming people, e.g. in medical diagnoses. 

Because DSS are explicitly designed to not automate but support decision-makers \cite{arnott2016critical}, the initially obvious idea emerges to address these problems by using DSS instead of fully automated systems. However, automation should not be interpreted as a binary state but instead as a continuum \cite{Parasuraman2000}. Negative impacts already occur at low automation levels \cite{Parasuraman2000}---as positive features of human decision-making are reduced such as human engagement. Therefore, when speaking about automated decision-making, we use the broader understanding of the continuum mentioned above, also including lower automation levels.
As many state-of-the-art DSS do include automated, AI-based recommendations \cite{turban2010decision}, they are subject to negative impacts, like automation bias in the short, reduced engagement in the medium, and deskilling in the long term. Thus, we perceive a major research gap in supporting human decision-making without those downsides, and formulate:

\textbf{RQ:} \textit{How can we design AI for decision support without introducing automation disadvantages?} 

%Conceptualization
Based on automation and DSS research, we conceptualize a new class of DSS, \textbf{\textit{Intelligent Decision Assistance}} (IDA), that reduces automation-induced disadvantages while still preserving decision support levels. 
From the automation literature, we draw the critical evaluation of potential disadvantages of automated decision-making and the awareness of a continuum between full automation and human agency \cite{Parasuraman2000}. From DSS literature, we use the concept of guidance \cite{morana2017review}. Part of guidance theory is the explainability of DSS \cite{morana2017review} which is a traditional topic of IS research \cite{meske2020explainable}. We discuss various combinations of automation levels and explainability and eventually follow the idea of informative guidance as a guidance that foregoes to provide explicit recommendations \cite{silver1991decisional}.
In line with this notion, we propose to withhold the AI's decision and let the human ``brainstorm'' together with the AI by providing techniques from the Explainable AI (XAI) knowledge base \cite{adadi2018peeking}, such as examples, counterfactuals, or feature importance. 
After conceptualizing IDA and deriving hypotheses on its impact, we provide first evidence for their validity through a systematic evaluation of empirical studies in the literature.
%Contributions
With our work, we contribute to research and practice by conceptualizing a new class of DSS---\textbf{\textit{Intelligent Decision Assistance}}.

\section{Literature Review} 
\label{sec:litrev}
%Intro to methodology of literature review
In general, IS are designed to support or automate human decision-making \cite{zuboff1985automate}.
These two purposes are traditionally analyzed in two different research streams:
\textit{decision support} is traditionally covered in DSS literature \cite{power2007brief}, while \textit{Automation} is mainly addressed in Ergonomics literature \cite{coombs2020strategic}.

\subsection{Decision Support Systems}
%Definition, Goals, History, Concepts
%Definition
DSS represent an important class of IS that aim to provide decisional advice \cite{arnott2016critical}.
In general, ``DSS is a content-free expression, which means that there is no universally accepted definition'' \cite[p. 16]{turban2010decision}. However, DSS can be used as an umbrella term to describe any computerized system that supports decision-making in an organization \cite{turban2010decision}. 
Originally, DSS were defined as supportive IT-based systems, aiming at supporting and improving managerial decision-making \cite{arnott2016critical,young1983right}. 
Later developments in DSS opened the area for application to all levels of an organization \cite{arnott2016critical}. 
%Goals
In contrast to other IS, DSS focuses on decision-making effectiveness and decision-making efficiency rather than efficiency alone \cite{evans1989assessing}. 

%Concepts
%Decision making
In general, the decision-making process consists of three phases that are supported through DSS---the intelligence, design, and choice phase \cite{simon1960new}. In the intelligence phase, the decision-maker searches, classifies and decomposes problems \cite[p. 48-49]{turban2010decision}.
In the design phase, decision alternatives are derived \cite[p. 50]{turban2010decision}. 
Finally, in the choice phase, the critical phase of decision making, the decisions are chosen \cite[p. 58]{turban2010decision}. 

%Decisional guidance 
An important concept of decision support is decisional guidance that has a long-lasting history in IS literature \cite{morana2017review}. 
\citet{silver2015decisional} differentiates in the form of guidance, which can be either suggestive, quasi-suggestive, or informative. 
Suggestive guidance makes judgmental recommendations that can also be a set of alliterative decisions \cite[p. 94]{silver2015decisional}.
Quasi-suggestive guidance is guidance ``that does not explicitly make a recommendation but from which one can directly infer a recommendation or direction'' \cite[p. 109]{silver2015decisional}. Lastly, informative guidance provides decision-makers only with decision-relevant information without suggesting or implying how to act. 

Another form of guidance is the explainability of the DSS \cite{morana2017review}. 
%Transition to XAI
Explainability is a concept with a long tradition in IS \cite{gregor1999explanations}.
With the rise of expert systems, knowledge-based systems, and intelligent agents in the 1980s and 1990s, the IS community has built the basis for research on explainability \cite{meske2020explainable}. 
In particular, the research stream of Explainable AI (XAI), which addresses the opaqueness of AI-based systems, is gaining momentum.
The term XAI was first coined by \citet{van2004explainable} to describe the ability of their system to explain the behavior of agents. The current rise of XAI is driven by the need to increase the interpretability of complex models \cite{wanner2020white}. In contrast to interpretable linear models, more elaborate models can achieve higher performance \cite{briscoe2011conceptual}. However, their inner workings are hard to grasp for humans.  XAI encompasses a wide spectrum of algorithms. These algorithms can be differentiated by their complexity, their scope, and their level of dependency \cite{adadi2018peeking}. The interpretability of a model directly depends on its complexity. \citet{wanner2020white} define three types of complexity---white, grey, and black-box models. They define white-box models as models with perfect transparency, such as linear regressions. These models do not need additional explainability techniques but are intrinsically explainable. Black-box models, like neural networks, on the other hand, tend to achieve higher performance but lack interpretability. Lastly, grey-box models are not inherently interpretable but are made interpretable with the help of additional explanation techniques. These techniques can be further differentiated in terms of their scope, i.e., being global or local explanations\cite{adadi2018peeking}: Global XAI techniques address holistic explanations of the models as a whole. In contrast, local explanations function on an individual instance basis. Besides the scope, XAI techniques can also be differentiated with regard to being model-specific or model agnostic.

\subsection{Automation}
%Definition, Goals, History, Concepts, AI
%Definition
Research on automation is an essential part of IS research \cite{frank1998increasing} and has been around for more than a century \cite{lacity2016new} with the overarching goal to increase the efficiency of work by using automation as a means \cite{hitomi1994automation}.
%Goals
In general, humans are performing worse than machines in conducting repetitive tasks and are influenced by cognitive bias \cite{heer2019agency}. Thereby, automation can reduce human bias-induced errors. 
Automated decision-making applications are designed to minimize human involvement and relieve humans from exhaustive tasks \cite{Harris2005}. Additionally, automation acts as an ``talent multiplier'' that scales human expertise and frees up human capacity to focus on more valuable work \cite{Harris2005}. 

%Concepts
%Levels of automation
Traditionally, automation has been seen as a binary state---either none or fully automatic \cite{endsley1999level}. However, \citet[p. 287]{Parasuraman2000} define automation as ``the full or \textit{partial} replacement of a function previously carried out by the human operator'' which implies that automation may occur on different levels. The authors propose a taxonomy of automation and develop ten levels. While humans are responsible for decision making at the first five levels, AI has control at the last five levels up to full autonomy at level ten. 

%Process
Beyond developing the 10-level taxonomy, \citet{Parasuraman2000} provide a four-stage model of automated human information processing consisting of information acquisition, information analysis, decision and action selection, and action implementation. This model allows to precisely specify which stage is automated in the decision process. 

%Discuss the drawbacks 
Although automation has many advantages, some authors have expressed challenges, such as  automation bias or cognitive skill reduction leading possibly to deskilling \cite{bainbridge1983ironies}. In the following, we discuss these disadvantages which essentially represent the problem with current approaches that we want to solve.

%Automation Bias
In the short-term, automation might lead to \textit{Automation bias} which is the “tendency to use automated cues as a heuristic replacement for vigilant information seeking and processing” \cite{Mosier1999}---essentially representing an over-reliance on AI recommendations. 
For this reason, sometimes high levels of automation are not desirable if the automation is not perfectly reliable and recommends wrong decisions \cite{sarter2001supporting}. These wrong recommendations then can lead to a negative switch from a previously correct human decision \cite{goddard2012automation}. 

Furthermore, in the long-term, automation bias can result in \textit{deskilling}, either because of the reduction of existing skills or due to the lack of skill development in general
\cite{meske2020explainable,sutton2018much}. This attacks the collective intellectual capital that is the key asset of many organizations \cite{asatiani2019implementation}.
Many factors might eventually result in deskilling. One factor is the reduced amount of stored information in memory, and more importantly, the reduced mental capability to store information,  when using automation, which is commonly known as the ``Google effect'' \cite{sparrow2011google}. Users seem to reduce investing energy into storing things that can be easily retrieved \cite{sutton2018much}. 
%This essentially boils down to a trade-off in the level of cognitive load. While traditionally, cognitive load is perceived as something solely negative, 
%Research in automation bias has discussed the tendency that humans always the path of least cognitive effort \cite{Mosier1999}.

Research shows that human \textit{engagement} in the task is particularly important to keep up the vigilantly \cite{Mosier1999}. Engagement is a psychological state that  is broadly defined as an ``individual’s  involvement  and satisfaction  as  well  as  enthusiasm  for  work'' \cite[p. 269]{harter2002business} that could reduce potential deskilling \cite{asatiani2019implementation}.
Exemplary, the danger of deskilling can be highlighted with an intelligent asset solution for financial markets.
Thereby, the engagement of the broker in the task will reduce which may lead finally to deskilling.
Therefore, within the company implementing that solution, the broker deskills---while brokers from companies not implementing the project stay skilled. 
In the long-term, the environment may eventually change, for example, because of new regulations. Therefore, existing AI solutions need to be built and trained. One of the most important factors in the development process is domain knowledge which may now be reduced due to deskilling. 
If other companies did not implement AI, they can build and adapt faster and will, therefore, have competitive advantages .

This long-term disadvantage of automated decision support leads to a discussion of efficiency in the short and long-term in human-AI systems. In the short-term AI might increase performance. However, in the long-term due to deskilling, AI systems will not be effectively further trained and evolved. This potentially results in severe negative long-term effects. %In the following section, we conceptualize a solution approach.

\section{Conceptualization of Intelligent Decision Assistance} 
\label{sec:conceptualization}
%Automation to characterize issue
%DSS to find solution
In this section, we use the previously depicted research streams of DSS and automation and synthesize them to conceptualize a solution against the disadvantages of automation.
Subsequently, we discuss three particular techniques of this concept

%Derivation level of agency and level of guidance
We see two main dimensions that influence the undesired effects of automation, which we discuss in more detail below: First, the general level of human control and agency \cite{Parasuraman2000} and, second, the form and degree of explainability \cite{morana2017review}. 

%Level of agency
Which level of human agency in automation should be implemented is a notorious discussion in automation literature \cite{heer2019agency}.
\citet{asatiani2019implementation} have discussed that retaining control of human workers may help to sustain their skill level. Similar, \citet{endsley1997role} argued that lower automation levels, in general, can keep them cognitively engaged. 

%Degree of explainability
Regarding the second dimension, the literature suggests ``that a seamless, collaborative interaction between human agents and automated tools, as opposed to using automation as an isolated ``black box'', could help to prevent the ill effects of deskilling'' \cite[p. 6]{asatiani2019implementation}. 
As discussed, the research stream of XAI addresses this ``black box'' issue in AI-based automated decision-making. Recent examples \cite{lai2019human,ribeiro2018anchors}
demonstrate the capability of XAI to support end-users in their decision-making. By varying the ``degree'' of explanations, i.e. the system's transparency \cite{vossing2020designing}, we believe different effects on the negative aspects of automation could be influenced. 
On the one hand, some might argue that more explainability is always better. However, the latest research suggests that a high level of automation paired with high explainability might just result in automation bias \cite{HemmerSchemmerVoessing2021_1000133631}. Furthermore, the degree of explainability should be adapted to the profession and experience of the end-user, e.g. novice users might need more intuitive and simpler explanations while data scientists can get the full degree of potential explanations \cite{kuehl2020you}.
These examples show that also the degree of explainability needs to be chosen thoughtfully. 

%Move on this two dimensions
%These two dimensions now allow us to discuss particular combinations. 

%Final derivation: informative guidance --> XAI without recommendation /DSS
As introduced, there are many forms of guidance---suggestive, quasi-suggestive, and informative guidance \cite{silver1991decisional}. 
Suggestive guidance provides the decision-maker with explicit recommendations and tries to increase the guidance of this recommendation. However, as \citet{Parasuraman2000} states, also partially automated systems can lead to automation bias and skill degradation. 
In contrast, as mentioned, informative decisional guidance is a form of guidance where users do not receive explicit recommendations \cite{silver1991decisional}. 
We follow this line of reasoning and propose a system could simply withhold its recommendation---although it is aware of that recommendation. \citet{parkes2012persuasive} validates that suggestive guidance---which is actually a form of automated decision making---can lead to automation bias, while informative guidance does not have such effects. Research also shows that the effects of the types of guidance vary depending on the task complexity. \citet{montazemi1996effectiveness} found that suggestive guidance is better for less complex tasks and informative guidance is better with increasing task complexity. This argument strengthens our derivation. Following this line of thought gives rise to the idea to set the degree of automation to almost zero and withhold explicit AI recommendations while keeping support through explanations up.
By doing so, we can minimize the drawbacks of automation while still assisting human decision-making. We are creating intelligent systems that are fully capable of solving issues on their own but use their capabilities to inspire and support instead of automating. 
%Definition
Based on the derivation, we name this new class of DSS, Intelligent Decision Assistance (IDA) and define it as follows: 

\textbf{Definition: }\textit{Intelligent Decision Assistance (IDA) is an AI that a) supports humans, b) does not recommend explicit decisions or actions, and c) explains its reasoning}

%Discussion (Benefits)
Referring to the three phases of decision making---intelligence, design, and choice---we mainly support with this approach the intelligence and to some extent the design phase. 
In terms of final effects on the human, we derive three hypotheses (engagement, performance, automation disadvantages).

First, IDA provides decision-makers with options to actively engage with the task by interactively requesting explanations, interpreting them and essentially communicating with the AI. %Literature shows that this active engagement will help to learn the underlying concepts of the task \cite{glover1997influence}. 
As \citet{asatiani2019implementation} \cite{asatiani2019implementation} have discussed providing explanations instead of using automation as an isolated ``blackbox'' could result in an engaged human-AI collaboration. Thus, we hypothesize:

\textbf{H1:} \textit{IDA increases engagement with the task.}

Beyond that, we hypothesize that IDA should increase human performance. While especially, if the automation is far better than the human, IDA will most likely not exceed automated decision-making, it should still improve the performance by providing guidance and especially insights.
Therefore, we formulate:

\textbf{H2:} \textit{IDA performance outperforms the human alone.}

Lastly, because IDA does not incorporate higher levels of automation it should reduce automation disadvantages and especially prevent deskilling. Therefore, we formulate the following hypothesis:

\textbf{H3:} \textit{IDA reduces automation induced disadvantages.}

In the next section, we are going to test these hypotheses based on empirical studies in the literature.

\begin{figure}[!htbp]
    \centering
	\includegraphics[width=\linewidth]{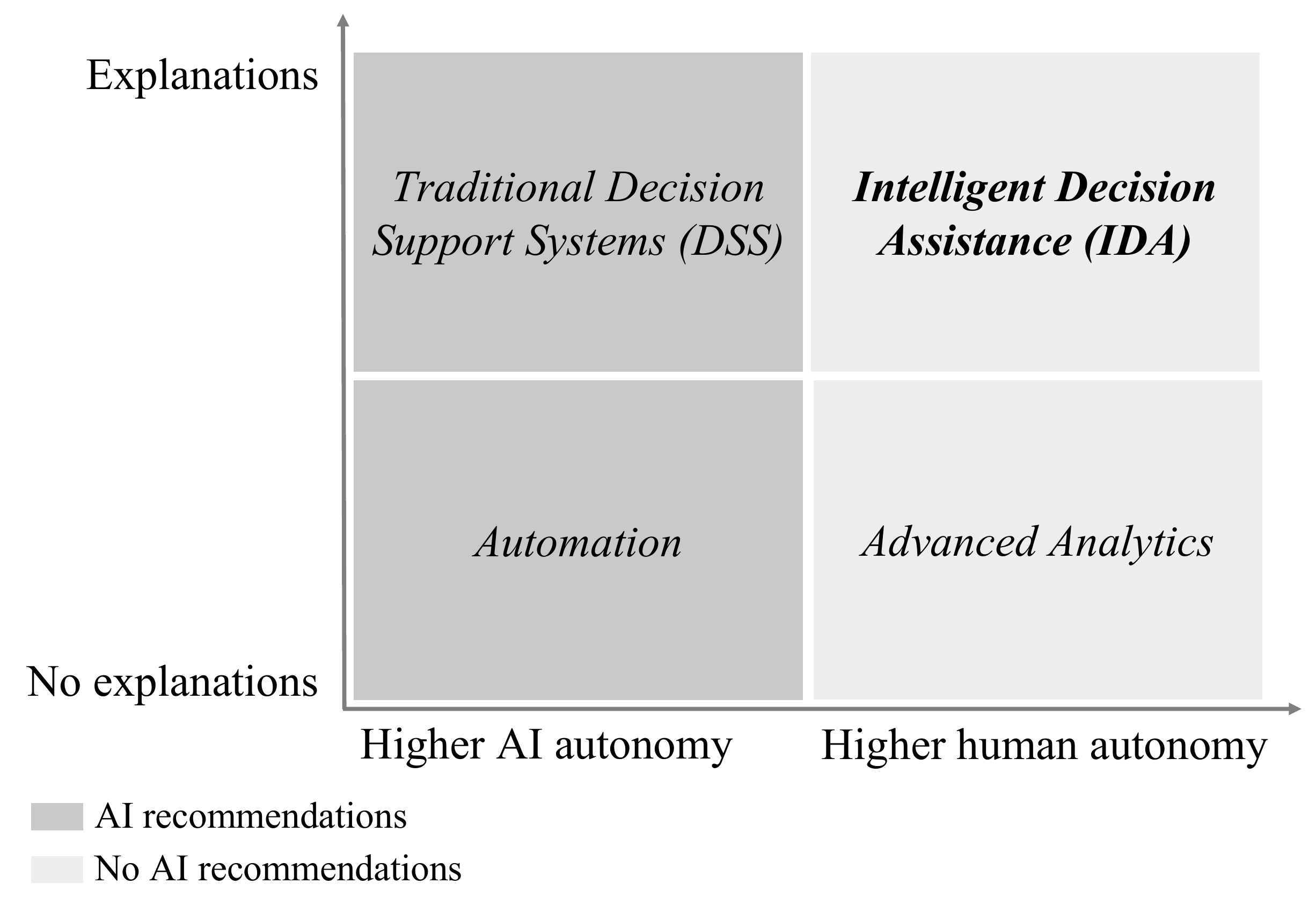}
	\caption{Positioning of Intelligent Decision Assistance on the two dimensions of explainability and degree of automation}
	\label{fig:dimensions}
\end{figure}

\begin{figure*}[!htbp]
    \centering
	\includegraphics[width=0.75\linewidth]{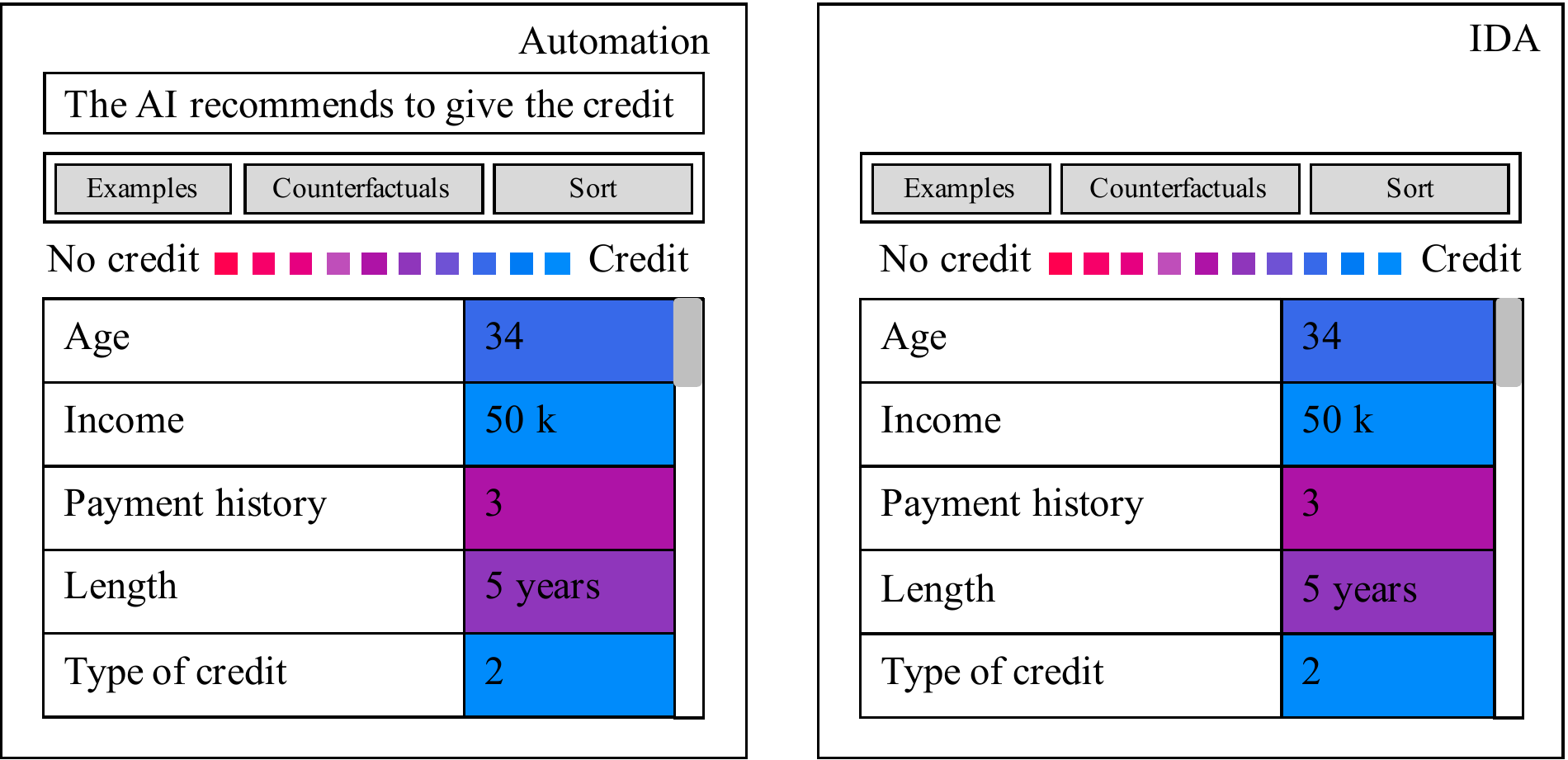}
	\caption{Comparison of traditional automated decision-making and Intelligent Decision Assistance (IDA)}
	\label{fig:IDA}
\end{figure*}

%Move on dimensions
\Cref{fig:dimensions} depicts IDA in the continuum of both discussed dimensions. We depict different types of systems for decision-making. At a high level of automation and almost no explanations, we position automation \cite{turban2010decision}.  Traditional DSS come also usually with a higher level of automation, through providing explicit recommendation, but additionally provide explanations for the decision-maker. 
%Delimitation
We delimit ourselves from DSS that use AI to transform unstructured data into structured data and DSS that use AI to produce a pre-decision output, e.g. a forecast. As stated, \citet{Parasuraman2000} define four stages of automation---information acquisition, information analysis, decision-making, and actions. Following this classification, we focus on the decision-making level. This classification allows us also to differentiate IDA from Advanced Analytics \cite{watson2014tutorial}. 
While advanced analytics may incorporate AI solutions they are always on the information acquisition or analysis level. In contrast, IDA allows the decision-maker to actively engage on the decision level and is positioned in the right top corner of \Cref{fig:dimensions} with high explainability and full human autonomy.

%\subsection{Explanation Techniques} %Discuss with loan approval example
Now that we derived, defined, and delimit IDA, we discuss specific explanation techniques that support IDA and consequently pose valid implementation options. Specifically, we discuss feature importance, example-based explanations, and counterfactual explanations. We explain these features based on the example of a loan approval decision-making task.

%Definition & Explanation & IDA application

\textbf{Feature importance}: 
Feature importance is a model-agnostic technique that gives the decision-maker information about the importance of specific data points. Two famous algorithms of feature importance are LIME \cite{lime} and SHAP \cite{NIPS2017_7062}. 
In a loan approval decision where the banker has information about past credits, expenses, demographics, etc., one could now train artificial intelligence to make this decision and recommend explicit decisions. 
In contrast, IDA would withhold the specific AI decision but provide the decision-maker, i.e. the banker, with information on which data was in particular important for the AI's decision. 
In an IDA this information could now be used for various use cases. Now in the time of big data, e.g. having many information on customers, one particular great use case would be to filter or sort the features in an intelligent way based on the feature importance.

\textbf{Example-based explanations}: 
%Definition
Example-based explanations provide historical data that is similar to the current instance \cite{van2021evaluating}.
%Explanation
Example-based explanations, therefore essentially represent some form of information retrieval.
%Motivation
Research in psychology states that humans prefer explanations that show examples \cite{cai2019effects}.
Furthermore, examples can be used within complex tasks \cite{glaser1986intelligence}.  
%Example 
Referring to our loan approval case, the decision-maker would receive information on past approvals that were similar.
%IDA
In an IDA, the decision-maker would get information about similar historical cases that are labeled. Based on these examples, the decision-maker should be able to infer differences or similarities. 

\textbf{Counterfactual explanation}: 
%Definition
Counterfactual explanations give information on what the smallest change would be to get a different AI decision \cite{wachter2017counterfactual}. 
%Explanation
Counterfactual explanations take a similar form to the statement \cite{schoeffer2021study}:
``You were denied a loan because your annual income was 20,000. If your income had been 45,000, you would have been offered a loan.''
%IDA
In an IDA a counterfactual explanation would look like the following: ``Your current annual income is £30,000. If your income would be £45,000, the AI's decision would change.''
This type of non-intrusive explanation would lead to an increased thought process of the decision-maker.

%Describe example
\Cref{fig:IDA} highlights the idea of IDA  for a credit allowance example. On the left side, we display a traditional interface for automated decision-making. On the right side, IDA is visualized. In the traditional interface, the decision-maker gets a specific recommendation. Additionally, the decision-maker gets the available information on the credit applicant, the importance of the features for the decision, and optional explanation options. In contrast, an IDA does not provide a specific recommendation, but rather  various XAI techniques that allow the decision-maker to ``brainstorm'' with the AI.

\section{Validation Study}
\label{sec:valstudy}
After deriving a conceptualization of IDA, we validate our concept by conducting a literature-based validation study based on the methodology outlined by \citet{brocke2009reconstructing}. 
The goal of the study is to find empirical studies that tested variations of automation and explainability and to analyze whether the findings do support our hypotheses above. This means they should address the degree of automation and explainability. 
For this reason, our search string consists of two main parts.
The first reflects XAI, including relevant synonyms, such as “explainable AI” or ``interpretability'' comprises of ``Artificial Intelligence''. The second part comprised synonyms of behavioral experiments, e.g., ``user study'' or ``user evaluation''. To find the synonyms, we initiated our SLR with an explorative search. The search string was iteratively extended resulting in the following final search string: 

\textit{TITLE-ABS-KEY(``explainable artificial intelligence'' OR XAI OR ``explainable AI'' OR ( ( interpretability OR explanation ) AND (``artificial intelligence'' OR ai OR ``machine learning'' ) ) ) AND ( ``human performance'' OR ``human accuracy'' OR ``user study'' OR ``empirical study'' OR ``online experiment'' OR ``human experiment'' OR ``behavioral experiment'' OR ``human evaluation'' OR ``user evaluation'')} 

Then, we selected an appropriate database. Our exploratory search indicated that relevant work is dispersed across multiple disciplines, publishers, conferences, and journals. For this reason, we chose the SCOPUS database, to ensure comprehensive coverage.
Following that, we defined our inclusion criteria. We included every article that (a) conducted empirical research, (b) reported performance measures,  (c) focused on an application context where AI supports humans on the decision level, and (d) provided an IDA setting.
With our search string defined, we conducted the SLR from January to March 2021. We identified 256 articles through the keyword-based search. As a next step, we analyzed the abstract of each article and filtered based on our inclusion criteria, leading to 61 articles. Afterward, two independent researchers read all articles in detail and applied the inclusion criteria again. Based on these, we conducted a forward and backward search. This led to a total of five articles that were consequently analyzed in-depth to collect data about each experiment. 
The data collection process was conducted by two independent researchers who discussed and homogenized differences. The main focus of the validation study was to extract the treatments and outcomes of each experiment reported in the studies. For example, if two XAI techniques were used and compared as separate experimental treatments we added two entries into our database. 
%Results
In total, we identified five articles and 12 experiments  \cite{carton2020feature,chu2020visual,lai2019human,lai2020chicago,schmidt2019quantifying}.
%Quantitative
In the following, we describe the studies and their results with regard to IDA in detail.

\citet{carton2020feature} conduct an experiment on online toxicity classification of social media posts. They use feature importance to highlight words that were relevant for the classification. As one condition they have the prediction presence. In their experiment, they find no significant effect of examples. However, they find signs of automation bias:``We find that the presence of a visible model prediction tends to bias subjects in favor of the prediction, whether it is correct or incorrect.'' \cite[p. 101]{carton2020feature}

\citet{chu2020visual} conduct an experiment on age guessing supported through AI. They test three different conditions of explanations and the visibility of AI predictions.
The authors found no significant effects of explanations but also signs of automation bias: ``The predictions generally help whenever the human is inaccurate [...], but can hurt when the human is accurate and the model is inaccurate [...].'' \cite[p. 5]{chu2020visual}

%Task, XAI, IDAs, Results
\citet{lai2019human} and \citet{lai2020chicago} refer in their studies also to the ten levels of automation introduced by \cite{Parasuraman2000} and test various XAI techniques without ever displaying what the actual AI’s decision is on a deception detection task. For example, they highlight all words that were relevant for the decision (unsigned) \cite{lai2020chicago}. Another condition was to colorize this highlight differently depending on the influence of the words (signed). Their results show that signed highlights result in a significant increase in XAI-assisted performance ($70.7\%$ for signed, and $60.4\%$ for human performance) \cite{lai2020chicago}. In \citet{lai2019human} they test additionally the influence of example-based explanations with also positive but not significant effects. However, also in \citet{lai2019human} two highlight-based conditions showed significant positive effects in terms of short-term performance.

Lastly, \citet{schmidt2019quantifying} conduct two different tasks in their experiment---a book category classification based on their descriptions and a movie rating classification. They test two different XAI algorithms, both feature importance techniques to highlight important words. Both data sets and both XAI algorithms show an increase in IDA performance with one algorithms generating significant results on both data sets.

%Add Table
\begin{table}[thb]
\centering
\caption{ Validation study results \vskip 3pt }
\label{tab:validation_study}
\resizebox{\columnwidth}{!}{%
\begin{tabular}{c|l|l|l}
\hline \bf Source &\bf Engagement &\bf Performance  &\bf Automation  \\ \hline
\cite{carton2020feature} & No Measurement & No effect & Automation Bias\\
\cite{chu2020visual} & No Measurement & No effect & Automation Bias \\
\cite{lai2019human} & No Measurement & Improvement & No Measurement\\
\cite{lai2020chicago} & No Measurement & Improvement& No Measurement\\
\cite{schmidt2019quantifying} & No Measurement & Improvement& No Measurement  \\
\hline
\end{tabular}}%
\end{table}

\Cref{tab:validation_study} summarizes our results of the validation study. Regarding our first hypothesis (\textbf{H1}), we can see that current research fails to provide insights into the effect of IDA on engagement.  Regarding \textbf{H2}, three papers validated our hypotheses that IDA performance should exceed human performance. Lastly, regarding \textbf{H3}, two of the studies showed signs of Automation Bias in the presence of explicit AI recommendations, which is an indicator of potential long-term deskilling effects \cite{meske2020explainable,sutton2018much}.

\section{Discussion}
\label{sec:discussion}
%Discuss validation study
Overall, the validation study provides first support for the hypotheses on the impact of IDA and highlights the potential of IDA through five experiments with significant positive effects and none with significant negative effects.
Furthermore, the study shows that current research lacks insights on the influence of IDA on engagement which should be addressed in future research.
%Discuss idea
%Some studies discuss disadvantages of automation, such as automation bias when providing explicit AI recommendations which highlights the potential benefits of IDA.% We can conclude, IDA allows for maximum support while not suffering from the disadvantages of automation. Furthermore, IDA can enhance the exploratory data analysis, diagnostic tasks, and in general creativity.

%Limitation of idea
IDA has of course also limitations. One of them might be the perceived usefulness. Telling the decision-maker that the AI would be theoretically capable of providing them with a recommendation but this recommendation is to withhold may be perceived as annoying for decision-makers, especially if they are under time pressure. Therefore, the advantages of IDA need to be highlighted. One attenuated option could be to show the explanations on default, but the recommendation just on request.  
Another limitation is the potential high computational costs. Some XAI techniques, e.g. SHAP values \cite{NIPS2017_7062}, are computational inefficient. Therefore, the computational costs, especially in comparison to traditional analytics tools might be much higher. This trade-off has to be determined for individual cases.

%when should it be used?
We want to clarify that IDA should not be applied in every use case. We explicitly derive this idea for knowledge work and not for repetitive structured work. Especially for jobs where the disadvantages of automation are critical, IDA should be taken into account. Among others, in high stake decision-making such as medicine, law, or human resource. But also in knowledge-intensive areas where the competitive advantage is based on knowledge, such as finance. 
However, as pointed out by \citet{endsley1999level}, for structured tasks that require low flexibility and have a high system performance, full automation can be the best option. %\Cref{fig:future} visualizes this ambiguity. 

%\begin{figure}[!htbp]
%    \centering
%	\includegraphics[width=0.9\linewidth]{images/future.pdf}
%	\caption{Past, present and future of automation, Decision %Support Systems (DSS) and Intelligent Decision Assistance (IDA)}
%	\label{fig:future}
%\end{figure}

Additionally, we want to discuss an additional advantage that may have a temporary influence on the adoption of IDA. Paragraph 22 of the GDPR states: ``The data subject shall have the right not to be subject to a decision based solely on automated processing [...].'' \cite{gdpr_2018}
This means that in some cases automated decision-making is simply forbidden. Here the best possible augmentation through IDAs could be a valuable approach.

%Fairness
Furthermore, IDAs could have a positive influence on the fairness of AI-enhanced decision-making. AI algorithms can have biases that can lead to unfair decision-making. With IDAs, we allow people to have full control over the final decision and can thus reduce bias. 

%Future work
%Open research questions
Finally, there are some open questions.
Future work should empirically validate whether IDAs prevent deskilling and other automation disadvantages and in contrast increases engagement. 
Furthermore, one should access the efficiency effects of IDA on human decision-making. For example, \citet{fazlollahi1995evaluation} find that decisional guidance increases decision time. However, also direct recommendations may decrease efficiency if they lead to cognitive dissonance and consequently to an in-depth analysis of the decision-maker. The efficiency of IDAs needs to be compared to pure human and automated approaches.

\section{Conclusion}
\label{sec:conclusion}
The main goal of this study was to conceptualize a solution to  automation-induced disadvantages, such as automation bias or deskilling. To do so, we initiated our research by conducting a literature review of automation and DSS literature. 
Based on these two research streams, we conceptualized a new class of DSS, namely \textbf{\textit{Intelligent Decision Assistance}} (IDA). IDA augments human decision-making through Explainable AI (XAI) while withholding explicit AI recommendations. Thereby, IDA aims to provide insight into the data without generating automation disadvantages. Subsequently, we validated our conceptualization by searching for empirical literature which shows first evidence of our hypotheses.

%Results of validation study & contributions
Our contributions are threefold: First, we synthesize the body of knowledge in automation sciences and decision support literature. Second, we conceptualize a new class of systems---IDA---and third, we test three hypotheses regarding the potential of IDA.

%Relevance for IS
Unleashing the potential of IDA requires a multidimensional design process.  For this reason, we see the IS research community as the predestined
research discipline to advance research in this field. We hope to motivate IS researchers and
practitioners to actively participate in the exploration of IDA.

% if added before the last page, this command can help balancing columns
%\addtolength{\textheight}{-.2cm} 
%Bibliography 
%\bibliographystyle{ieeetr}
\bibliographystyle{unsrtnat}
\bibliography{sample}

\end{document}